\documentclass{article}%
\usepackage{amsfonts}
\usepackage{amsmath}
\usepackage{amssymb}
\usepackage{graphicx}%
\providecommand{\U}[1]{\protect\rule{.1in}{.1in}}
\newtheorem{theorem}{Theorem}

\newtheorem{conclusion}[theorem]{Conclusion}

\newtheorem{conjecture}[theorem]{Conjecture}

\newtheorem{proposition}[theorem]{Proposition}

\begin{document}

\title{Algebraic symplectic reduction and quantization of singular spaces }
\author{V. P. Palamodov}
\date{}
\maketitle

\textbf{Abstract.} The algebraic method of singular reduction is applied for
non regular group action on manifolds which provides singular Poisson spaces.
For some examples of singular Poisson spaces the deformation quantization is
explicitly constructed. It is shown that for the flat phase space with the
classical moment map and the orthogonal group action the deformation
quantization converges for the class of entire functions.

Non free


\section{Introduction}

The problem of systems with constraints in the quantum field theory comes to
Dirac \cite{D}. The general method of Meyer-Marsden-Weinstein provides the
reduction of a symplectic manifold with constraint and a free group action. If
the group action is not free and the constraint locus is singular. The
singular points are often the most interesting because they have smaller
orbits and larger symmetry. Sniatycki\ and Weinstein \cite{SW} applied a pure
algebraic method for symplectic reduction in a modelling case. The problem of
singular symplectic reduction of the angular momentum was studied in
\cite{BG}, \cite{AGJ} by geometric methods. Batalin-Vilkovisky-Fradkin's
method \cite{FV}, \cite{BV} was proposed for gauge systems. In \cite{BHP} the
BRST\ method was developed based on the rather complicated homological
construction including ghosts fields.

The method of algebraic singular reduction can be applied to any algebraic
Poisson manifold $\left(  X,q\right)  $ with an algebraic momentum map and
action of an algebraic group $G$. It ends up to an affine Poisson algebraic
variety $\left(  X_{\mathrm{red}},q_{\mathrm{red}}\right)  $ with the algebra
sheaf $\mathcal{O}_{\mathrm{red}}$\ of $G$ invariant functions restricted to
the constraint locus. This variety is singular if the group action is not
free. This is the case of the Yang-Mills theory and general relativity where
the constraint locus\ has quadratic singularities and the reduced space
$X_{\mathrm{red}}$ is singular \cite{AMM}. We give here construction of
deformation quantization of some singular spaces $X_{\mathrm{red}}$. Other
examples are some singular K3 surfaces. Our method is based on the
Gr\"{o}newold-Moyal formula. In the simplest cases the flat phase space
associative product locally converges for entire holomorphic arguments.

The problem of quantization of spaces with singularities was rased by
Kontsevich \cite{Ko1}. To my best knowledge there is no examples of
deformation quantization of singular spaces so far. See \cite{Ste} for a
survey on quantization deformation and \cite{P1},\cite{P2} for basics of the
theory of associative deformations of singular spaces.

\section{Singular reduction}

We use the construction of singular reduction which is close to that of
\cite{SW}. Let $X$ be a real (or complex) algebraic variety endowed with a
Poisson bracket$\ q$ defined on the algebra of\ rational real or complex
functions on $X.$ In a more general setting let $\left(  X,\mathcal{O}%
_{X}\right)  $ be a real algebraic scheme with a Poisson biderivation
$q:\mathcal{O}_{X}\times\mathcal{O}_{X}\rightarrow\mathcal{O}_{X}$. An
algebraic group $G$ is defined on $X$ such that the bracket $q$ is $G$
covariant. Let $\mathcal{O}_{X/G}$ be the subsheaf of $\mathcal{O}_{X}$ of $G$
invariant germs. It is a sheaf of algebras defined on the space of orbits
$X/G$. The invariant Poisson bracket $q$\ can be lifted to a Poisson bracket
$q_{G}$ on $X/G.$

Let $J:X\rightarrow\mathfrak{g}^{\ast}$ be an algebraic momentum map, where
$\mathfrak{g}^{\ast}$ is the dual space to the Lie algebra $\mathfrak{g}$ of
$G$. The set $Y=J^{-1}\left(  0\right)  $ is a subscheme of $\mathcal{O}_{X}$
(called constraint locus) with the structure sheaf $\mathcal{O}_{Y}%
=\mathcal{O}_{X}/\left(  J\right)  ,$ where $\left(  J\right)  $ denotes the
ideal in $\mathcal{O}_{X}$ generated by the coordinates of $J$. We suppose
that $J$ is equivariant that is $J\left(  gx\right)  =\mathrm{Ad}gJ\left(
g^{-1}x\right)  $ for $g\in G.$ It follows that $(J)$ is $G\ $invariant and
$J$ can be lifted to a mapping $J_{G}$ defined on $Y/G$ making the diagram commutative:%

\[%
\begin{array}
[c]{ccccc}%
Y & \rightarrow & X & \overset{J}{\rightarrow} & \mathfrak{g}^{\ast}\\
\downarrow &  & \downarrow &  & \parallel\\
X_{\mathrm{red}}=Y/G & \rightarrow & X/G & \overset{J_{G}}{\rightarrow} &
\mathfrak{g}^{\ast}%
\end{array}
.
\]
We assume further that the action is hamiltonian that is for any
$v\in\mathfrak{g}$ and any $a\in\mathcal{O}_{X}$, we have%
\begin{equation}
q\left(  \left\langle v,J\right\rangle ,a\right)  =\mathrm{d}_{G}A\left(
v\right)  \left(  a\right)  \label{10}%
\end{equation}
where$\ A:X\times G\rightarrow X$ denotes the group action and $\mathrm{d}%
_{G}A:\mathfrak{g\rightarrow\Gamma}\left(  T\left(  X\right)  \right)  $ is
the tangent map.

\begin{proposition}
\label{red}The bracket $q$ can be lifted to a biderivation $q_{\mathrm{red}}$
on $X_{\mathrm{red}}\doteqdot Y/G.$ This is a Poisson bracket.
\end{proposition}

\textit{Proof. }Check that inclusion $q\left(  j,b\right)  \in\left(
J\right)  $ holds for any $j\in\left(  J\right)  $ and arbitrary
$b\in\mathcal{O}_{X/G}.$ Let $j=\left\langle v,J\right\rangle a$ for some
$a\in\mathcal{O}_{X}$ and $v\in\mathfrak{g.}$ We have%
\[
q\left(  j,b\right)  =\left\langle v,J\right\rangle q\left(  a,b\right)
+aq\left(  \left\langle v,J\right\rangle ,b\right)
\]
because $q$ is biderivation. The first term belongs to $(J)$ and by (\ref{10})%
\[
q\left(  \left\langle v,J\right\rangle ,b\right)  =\mathrm{d}_{G}A\left(
v\right)  \left(  b\right)  =0
\]
since $b$ is constant on any orbit and the field $\mathrm{d}_{G}A\left(
v\right)  $ is tangent to orbits of $G.$ Finally $q\left(  j,b\right)
\in\left(  J\right)  .\blacktriangleright$

The Poisson variety $\left(  X_{\mathrm{red}},\mathcal{O}_{Y/G}%
,q_{\mathrm{red}}\right)  $ will be called singular symplectic reduction of
$\left(  X,q,G,J\right)  $. This construction is translated to the category of
sheaves of smooth functions on $X$ with obvious modifications.

\section{Poisson bracket from hamiltonian fields}

Let$\ \mathcal{A}$ be a unitary commutative algebra over a field $\mathbb{K}$
of zero characteristic.

\begin{proposition}
\label{I}Let $q$ be a Poisson bracket on $\mathcal{A}.$ If $q\left(  q\left(
a,b\right)  ,\cdot\right)  =\mathtt{0\ }$for some $a,b\in\mathcal{A}$, then
the hamiltonian fields \textrm{A}$\left(  \cdot\right)  =q\left(
\cdot,a\right)  $ and \textrm{B}$\left(  \cdot\right)  =q\left(
b,\cdot\right)  $ commute.
\end{proposition}

This follows from the Jacobi identity. $\blacktriangleright$

For derivations \textrm{A}$,\ $\textrm{B} on $\mathcal{A},$ we define the
biderivation $\left(  \mathrm{A}\wedge\mathrm{B}\right)  \left(  a,b\right)
=\mathrm{A}\left(  a\right)  $\textrm{B}$\left(  b\right)  -\mathrm{B}\left(
a\right)  $\textrm{A}$\left(  b\right)  ,$ $a,b\in\mathcal{A}.$ For a
biderivation $q,$ we denote
\[
\mathrm{Jac}\left[  q\right]  \left(  a,b,c\right)  \equiv q\left(  q\left(
a,b\right)  ,c\right)  +q\left(  q\left(  b,c\right)  ,a\right)  +q\left(
q\left(  c,a\right)  ,b\right)
\]
and have $\mathrm{Jac}\left[  q\right]  =0$ if $q$ is the Poisson bracket.

\begin{proposition}
If \textrm{A}$_{i},$\textrm{B}$_{j},$ $i,j=1,..,n$ are commuting fields on
$\mathcal{A}$ then the bracket $q=\sum\mathrm{A}_{i}\wedge\mathrm{B}_{i}$
satisfies the Jacobi identity.
\end{proposition}

\textit{Proof.}\textrm{ }For $n=1$ this identity can be checked by a direct
computation. In the general case, we set $U=\sum t^{i}$\textrm{A}$_{i},V=\sum
t^{n-i}$\textrm{B}$_{i}$ where $t$ is a real parameter. The field $U$ and $V$
commute, hence $\mathrm{Jac}\left[  U\wedge V\right]  =0.$ The left hand side
is a polynomial in $t$ which vanishes identically. In particular the term with
$t^{n}$ vanishes which implies the statement. $\blacktriangleright$

We say that a subalgebra $\mathcal{B}$ of $\mathcal{A}$ is dense, if any
derivation $\delta:\mathcal{A\rightarrow A}$ such that $\delta\mid
\mathcal{B}=0$ vanishes on $\mathcal{A}.\ $

\begin{proposition}
\label{2n}Let $q$ be a Poisson bracket defined on $\mathcal{A}$. If there
exist elements $\mathrm{a}_{i},\mathrm{b}_{i}\in\mathcal{A}$,$\ i=1,..,n$ such
that
\begin{equation}
q\left(  \mathrm{a}_{i}\mathrm{,a}_{j}\right)  =q\left(  \mathrm{b}%
_{i}\mathrm{,b}_{j}\right)  =0=q\left(  \mathrm{a}_{i}\mathrm{,b}_{j}\right)
=0,i\neq j,\ q\left(  \mathrm{a}_{i}\mathrm{,b}_{i}\right)  =1,\ i=1,...,n
\label{2}%
\end{equation}
and $\mathrm{a}_{i},\mathrm{b}_{i}$ generate the dense subalgebra
$\mathcal{B}$ of$\ \mathcal{A}$ then
\begin{equation}
q\left(  \cdot,\cdot\right)  =\sum_{1}^{n}q\left(  \cdot,\mathrm{b}%
_{k}\right)  \wedge q\left(  \mathrm{a}_{k},\cdot\right)  . \label{1}%
\end{equation}

\end{proposition}

\textit{Proof.\ }Proposition \ref{I} implies commutativity of any pair of the
fields $q\left(  \cdot,\mathrm{b}_{k}\right)  ,\ q\left(  \mathrm{a}_{k}%
,\cdot\right)  $, $i,j=1,2,...,n.$ By (\ref{2}) the biderivation%
\[
\left[  \cdot,\cdot\right]  =\sum_{1}^{n}q\left(  \cdot,\mathrm{b}_{k}\right)
\wedge q\left(  \mathrm{a}_{k},\cdot\right)  \doteqdot\sum_{1}^{n}q\left(
\cdot,\mathrm{b}_{k}\right)  q\left(  \mathrm{a}_{k},\cdot\right)  -\sum
_{1}^{n}q\left(  \cdot,\mathrm{a}_{k}\right)  q\left(  \mathrm{b}_{k}%
,\cdot\right)  .
\]
fulfils
\[
\left[  \mathrm{a}_{i},\mathrm{b}_{j}\right]  =\sum_{k=1}^{n}q\left(
\mathrm{a}_{i},\mathrm{b}_{k}\right)  q\left(  \mathrm{a}_{k},\mathrm{b}%
_{j}\right)  -\sum_{k=1}^{n}q\left(  \mathrm{b}_{j},\mathrm{b}_{k}\right)
q\left(  \mathrm{a}_{k},\mathrm{a}_{i}\right)  =\delta_{ij}q\left(
\mathrm{a}_{i}\mathrm{,b}_{j}\right)
\]
that is $\left[  \mathrm{a}_{i},\mathrm{b}_{j}\right]  =q\left(
\mathrm{a}_{i},\mathrm{b}_{j}\right)  .$ Therefore $\left[  \mathrm{P,Q}%
\right]  =q\left(  \mathrm{P,Q}\right)  $ for any polynomials $\mathrm{P,Q}%
\in\mathcal{B}.$ The subalgebra $\mathcal{B}$ is dense in $\mathcal{A\ }$by
the\ assumption. This\ implies that the brackets coincide on $\mathcal{A}%
.\ \blacktriangleright$

\section{The Gr\"{o}newold-Moyal star product}

\begin{theorem}
\label{WM}For any Poisson bracket $q$ on $\mathcal{A}$ and any elements
$\mathrm{a}_{i},\mathrm{b}_{j}$ as in Proposition \ref{2n}, the
Gr\"{o}newold-Moyal (GM) product%
\begin{equation}
\left(  f\ast g\right)  \left(  t\right)  =fg+\sum_{k=1}^{\infty}\frac{t^{k}%
}{k!}\mathrm{Q}_{k}\left(  f,g\right)  \label{M}%
\end{equation}
defined on $\mathcal{A}$ is a deformation quantization of this bracket
where$\ \mathrm{Q}_{1}=q$ and for any$\ k=2,3,...,$%
\begin{align*}
\mathrm{Q}_{k}\left(  f,g\right)   &  =\sum_{j=0}^{k}\left(  -1\right)
^{j}\frac{k!}{j!\left(  k-j\right)  !}\sum_{i_{l}=1}^{n}\ \mathrm{A}_{i_{1}%
}...\mathrm{A}_{i_{j}}\mathrm{B}_{i_{j+1}}...\mathrm{B}_{i_{k}}\left(
f\right)  \ \mathrm{B}_{i_{1}}...\mathrm{B}_{i_{j}}\mathrm{A}_{i_{j+1}%
}...\mathrm{A}_{i_{k}}\left(  g\right)  ,\\
\mathrm{A}_{k}  &  =q\left(  \cdot,\mathrm{b}_{k}\right)  ,\ \mathrm{B}%
_{k}=q\left(  \mathrm{a}_{k},\cdot\right)  ,\ k=1,...,n.
\end{align*}

\end{theorem}

\textbf{Proof.} The fields $\mathrm{A}_{i},\mathrm{B}_{j}$ commute for
$i,j=1,...,n$ since of the Jacobi identity and (\ref{1})\ coincides with
(\ref{P}) for $\mathrm{a}_{i}=x^{i},$ $\mathrm{b}_{i}=\xi_{i}.$ Therefore
(\ref{M}) is the associative product which has the same form as the classical
Gr\"{o}newold-Moyal series. $\blacktriangleright$

\section{Invariant quantization of a flat phase space}

The phase space $T^{\ast}\left(  \mathbb{R}^{n}\right)  =\mathbb{R}^{n}%
\times\mathbb{R}^{n}$ is supplied with the Poisson bracket%
\begin{equation}
q\left(  a,b\right)  =\sum\frac{\partial a}{\partial x^{i}}\frac{\partial
b}{\partial\xi_{i}}-\frac{\partial a}{\partial\xi_{i}}\frac{\partial
b}{\partial x^{i}}, \label{P}%
\end{equation}
and the classical momentum map
\[
J:\mathbb{R}^{n}\times\mathbb{R}^{n}\rightarrow\wedge^{2}\mathbb{R}%
^{n},\ J\left(  x,\xi\right)  =x\wedge\xi.
\]
The action of the orthogonal group $\mathbf{O}\left(  n\right)  $ on
$\mathbb{R}^{n}\times\mathbb{R}^{n}:$ $\left(  x,\xi\right)  \mapsto\left(
Ux,U\xi\right)  $ preserves the Poisson bracket and $J$ is equivariant. The
constraint locus $Y=J^{-1}\left(  0\right)  $ consists of pairs $\left(
x,\xi\right)  $ of proportional vectors $x$ and $\xi$. For elements
$e_{jk}=y^{j}\partial/\partial y^{k}-y^{k}\partial/\partial y^{j},$ $j\neq
k=1,...,n$ of the Lie algebra of the group $\mathbf{O}\left(  n\right)  $, we
have $\left\langle e_{jk},J\right\rangle =x^{j}\xi_{k}-x^{k}\xi_{j}$ and
equation
\[
q\left(  \left\langle e_{jk},J\right\rangle ,a\right)  =\xi_{k}\frac{\partial
a}{\partial\xi_{j}}-\xi_{j}\frac{\partial a}{\partial\xi_{k}}+x^{j}%
\frac{\partial a}{\partial x^{k}}-x^{k}\frac{\partial a}{\partial x^{j}%
}=\mathrm{d}_{G}A\left(  e_{jk}\right)  \left(  a\right)
\]
implies (\ref{10}). By Proposition \ref{red} the bracket $q$ is lifted to the
Poisson bracket $q_{\mathrm{red}}$ in $Y/G$.

Let $\mathcal{A}$ be the algebra of real polynomials on $\mathbb{R}^{n}.$ The
algebra $\mathcal{A}_{X/G}$ of polynomials on $X=\mathbb{R}^{n}\times
\mathbb{R}^{n}$ invariant with respect to the action of is generated by
\[
s_{1}=\left\vert x\right\vert ^{2},\ s_{2}=\left\vert \xi\right\vert
^{2},\ s_{3}=\left\langle x,\xi\right\rangle .
\]
The restrictions of the generators on $Y$ fulfil equation $f\left(  s\right)
\doteqdot s_{3}^{2}-s_{1}s_{2}=0,\ $which implies $\mathcal{A}_{Y/G}%
\cong\mathbb{R}\left[  s_{1},s_{2},s_{3}\right]  /\left(  f\right)  $ and we
have
\[
q\left(  s_{1},s_{2}\right)  =4s_{3},\ q\left(  s_{1},s_{3}\right)
=2s_{1},\ q\left(  s_{2},s_{3}\right)  =-2s_{2}%
\]
or equivalently
\begin{equation}
q_{\mathrm{red}}=4s_{3}\partial_{1}\wedge\partial_{2}-2s_{2}\partial_{2}%
\wedge\partial_{3}-2s_{1}\partial_{3}\wedge\partial_{1}. \label{11}%
\end{equation}
The elements $\mathrm{a}_{1}=\sqrt{s_{1}},\ \mathrm{b}_{1}=\sqrt{s_{2}}$
belong to the quadratic extension $\mathcal{A}^{\ast}\mathcal{\ }$of the
polynomial algebra $\mathcal{A}_{Y/G}$. We have%
\[
q\left(  \mathrm{a}_{1},\mathrm{b}_{1}\right)  =\frac{4s_{3}}{2\sqrt{s_{1}%
}2\sqrt{s_{2}}}=1.
\]
Therefore the elements fulfil conditions of Proposition \ref{2n} for $n=1.$ It
follows that the bracket $q_{\mathrm{red}}$ admits the quantization of GM type
on the algebra $\mathcal{A}^{\ast}.$ The algebra $\mathcal{B}$ of polynomials
of \textrm{a}$_{1}$ and $\mathrm{b}_{1}$ is dense in $\mathcal{A}^{\ast}.$

\section{Convergence of the Gr\"{o}newold-Moyal series}

\begin{theorem}
The terms $\mathrm{Q}_{m}$ of the GM quantization of (\ref{11}) are
bidifferential operators with polynomial coefficients of degree $\leq m$ in
each argument.
\end{theorem}

\textit{Proof. }Denote $\mathrm{A}=q\left(  \cdot,\mathrm{b}_{1}\right)
,\ \mathrm{B}=q\left(  \mathrm{a}_{1},0\right)  .$ The bracket (\ref{11})\ has
linear coefficients. For an arbitrary even $k,$ we can write%

\begin{align}
\mathrm{Q}_{k}\left(  a,b\right)   &  =\sum_{i+j=k/2}\frac{k!}{2i!2j!}%
\mathrm{A}^{2i}\mathrm{B}^{2j}\left(  a\right)  \mathrm{B}^{2i}\mathrm{A}%
^{2j}\left(  b\right) \nonumber\\
&  -\sum_{i+j+1=k/2}\frac{k!}{\left(  2i+1\right)  !\left(  2j+1\right)
!}\mathrm{ABA}^{2i}\mathrm{B}^{2j}\left(  a\right)  \cdot\mathrm{ABA}%
^{2j}\mathrm{B}^{2i}\left(  b\right)  \label{8}%
\end{align}
since the fields
\begin{align*}
\mathrm{A}  &  =q\left(  \cdot,\sqrt{s_{2}}\right)  =2\sqrt{s_{1}}\partial
_{1}+\sqrt{s_{2}}\partial_{3},\ \\
\mathrm{B}  &  =q\left(  \cdot,\sqrt{s_{1}}\right)  =2\sqrt{s_{2}}\partial
_{2}+\sqrt{s_{1}}\partial_{3}%
\end{align*}
vanish on $f,$ commute and $\mathrm{A}\wedge\mathrm{B}=q.$ The bidifferential
operators are composed from the operators%
\begin{align*}
\mathrm{A}^{2}  &  =4s_{1}\partial_{1}^{2}+2\partial_{1}+s_{2}\partial_{3}%
^{2},\ \mathrm{B}^{2}=4s_{2}\partial_{2}^{2}+2\partial_{2}+s_{1}\partial
_{3},\\
\mathrm{BA}  &  =\mathrm{AB}=4s_{3}\partial_{1}\partial_{2}+2s_{1}\partial
_{1}\partial_{3}+2s_{2}\partial_{2}\partial_{3}+s_{3}\partial_{3}^{2}%
\end{align*}
which are second order differential operators with linear coefficients. For
odd $k$,%
\[
\mathrm{Q}_{k}\left(  a,b\right)  =\sum_{i+j=k-1}\left(  -1\right)  ^{j}%
\frac{\left(  k-1\right)  !}{i!j!}q\left(  \left(  \mathrm{A}^{i}%
\mathrm{B}^{j}\left(  a\right)  \right)  ,\mathrm{A}^{j}\mathrm{B}^{i}\left(
b\right)  \right)  .
\]

\begin{theorem}
For arbitrary holomorphic functions $a,b$ on $\mathbb{C}^{3}$ of exponential
type $\sigma,$ the GM series for the Poisson bracket (\ref{11}) converges for
$s$ in the ball $\left\{  \left\vert s\right\vert <1/4\sigma\right\}  $ for
$t$ satisfying $\left\vert t\right\vert <1/9\sigma^{1/2}.$
\end{theorem}

\textit{Proof. }Denote%
\[
\left\Vert a_{m}\right\Vert =\max_{\left\vert s\right\vert =1}\left\vert
a_{m}\left(  s\right)  \right\vert
\]
for a polynomial $a_{m}$ of degree $m$. Note that $\left\Vert \partial
_{i}a_{m}\right\Vert \leq m\left\Vert a_{m}\right\Vert ,\left\Vert s_{i}%
a_{m}\right\Vert =\left\Vert a_{m}\right\Vert ,\ i=1,...,n.$ If
\[
p\left(  s,D\right)  =\sum p_{ijk}s_{i}\partial_{j}\partial_{k},
\]
then degree of the polynomial $p\left(  s,D\right)  a_{m}$ is equal to $m-1$
and
\[
\left\Vert p\left(  s,D\right)  a_{m}\right\Vert \leq\left\vert s\right\vert
^{m}\frac{m!}{\left(  m-2\right)  !}\left\Vert p\right\Vert \left\Vert
a_{m}\right\Vert ,\ \left\Vert p\right\Vert =\sum\left\vert p_{ijk}\right\vert
.
\]
For any $i,j,$ $\mathrm{A}^{2i}\mathrm{B}^{2j}\left(  a_{m}\right)  $ is a
polynomial of degree $m-i-j$ and
\begin{align*}
\left\Vert \mathrm{AB}\left(  a_{m}\right)  \right\Vert  &  \leq9^{2}%
\frac{m!\left(  m-1\right)  !}{\left(  m-2\right)  !\left(  m-3\right)
!}\left\Vert a_{m}\right\Vert \leq9^{2}\left(  \frac{m!}{\left(  m-2\right)
!}\right)  ^{2}\left\Vert a_{m}\right\Vert \\
\left\Vert \mathrm{A}^{i}\mathrm{B}^{j}\left(  a_{m}\right)  \right\Vert  &
\leq9^{i+j}\frac{m!\left(  m-1\right)  !}{\left(  m-i-j\right)  !\left(
m-i-j-1\right)  !}\left\Vert a_{m}\right\Vert \\
&  \leq9^{i+j}\left(  \frac{m!}{\left(  m-i-j\right)  !}\right)
^{2}\left\Vert a_{m}\right\Vert ,\ i+j\geq2
\end{align*}
since
\[
\max\left(  \left\Vert \mathrm{AB}\right\Vert ,\left\Vert \mathrm{A}%
^{2}\right\Vert ,\left\Vert \mathrm{B}^{2}\right\Vert ,\left\Vert q\right\Vert
\right)  \leq9.
\]
It follows that for an arbitrary homogeneous polynomial $b_{n}$ of
degree\textrm{\ }$n,$ and any even $k,$%
\begin{align*}
\left\vert \mathrm{Q}_{k}\left(  a_{m},b_{n}\right)  \right\vert  &  \leq
9^{k}\sum_{i+j=k}\frac{k!}{i!j!}\left(  \frac{m!}{\left(  m-k/2\right)
!}\frac{n!}{\left(  n-k/2\right)  !}\right)  ^{2}\left\Vert a_{m}\right\Vert
\left\Vert b_{n}\right\Vert \left\vert s\right\vert ^{m+n-k/2}\\
&  \leq\left(  18\right)  ^{k}\left\vert s\right\vert ^{-k/2}\left(
k/2!\right)  ^{4}4^{m+n}\left\vert s\right\vert ^{m+n}\left\Vert
a_{m}\right\Vert \left\Vert b_{n}\right\Vert \\
&  \leq C9^{k}\left\vert s\right\vert ^{-k/2}k!\left(  k/2!\right)
^{2}\left(  4\left\vert s\right\vert \right)  ^{m+n}\left\Vert a_{m}%
\right\Vert \left\Vert b_{n}\right\Vert
\end{align*}
since $\left(  k/2!\right)  ^{2}\leq2\pi^{1/2}2^{-k}k!.$ Note that
$\mathrm{Q}_{k}\left(  a,b\right)  =0$ if $k/2>\mathrm{\min}\left\{
m,n\right\}  $. The similar estimate holds for any odd $k.$ Let%
\[
a=\sum a_{m},\ b=\sum b_{n}%
\]
for some series of homogeneous polynomials $a_{m},b_{n}.$ By the condition
both series fulfil%
\begin{equation}
\left\Vert a_{m}\right\Vert \leq C_{\varepsilon}\varepsilon^{m}m!,\ \left\Vert
b_{n}\right\Vert \leq C_{\varepsilon}\varepsilon^{n}n! \label{12}%
\end{equation}
for arbitrary $\varepsilon>\sigma$ and some constant $C_{\varepsilon}$ that
does not depend on $m$ and $n.$ For arbitrary polynomials $a_{m},$ and $b_{n}$
satisfying (\ref{12}), we finally obtain the inequality for $\left\vert
s\right\vert <1/4\varepsilon$ and $\left\vert t\right\vert <\varepsilon
^{-1/2}/9:$
\begin{align*}
\sum_{k}\frac{t^{k}}{k!}\left\vert \mathrm{Q}_{k}\left(  a,b\right)  \left(
s\right)  \right\vert  &  \leq C_{\varepsilon}^{\prime}\sum_{k}\left(
9\left\vert t\right\vert \right)  ^{k}\left\vert s\right\vert ^{-k/2}%
\sum_{\mathrm{\min}\left\{  m,n\right\}  >k/2}\left(  4\varepsilon\left\vert
s\right\vert \right)  ^{m+n}\frac{\left(  k/2!\right)  ^{2}}{m!n!}\\
&  \leq\frac{C_{\varepsilon}^{\prime}}{1-4\varepsilon\left\vert s\right\vert
}\sum_{k\geq0}\left(  9\left\vert t\right\vert \left\vert s\right\vert
^{-1/2}\right)  ^{k}\sum\left(  4\varepsilon\left\vert s\right\vert \right)
^{k/2}\\
&  =\frac{C_{\varepsilon}^{\prime}}{\left(  1-4\varepsilon\left\vert
s\right\vert \right)  \left(  1-9\varepsilon^{1/2}\left\vert t\right\vert
\right)  }.
\end{align*}
It follows the the series converges for any $s$ and $t$ such that $\left\vert
s\right\vert <1/4\sigma$ and $\left\vert t\right\vert <1/9\sigma^{1/2}$.
$\blacktriangleright$

\section{Commuting matrices}

Let $M_{2}$ be the space of $2\times2$-matrices with complex entries. The
manifold $X=M_{2}\times M_{2}$ is endowed with the Poisson bracket%
\begin{equation}
q=\sum_{k=1}^{4}\frac{\partial}{\partial a_{k}}\wedge\frac{\partial}{\partial
b_{k}}, \label{9}%
\end{equation}
where
\[
\mathsf{A}=\left(
\begin{array}
[c]{cc}%
a_{1} & a_{3}\\
a_{4} & a_{2}%
\end{array}
\right)  ,\ \mathsf{B}=\left(
\begin{array}
[c]{cc}%
b_{1} & b_{3}\\
b_{4} & b_{2}%
\end{array}
\right)
\]
are coordinates in $X.$ The group $\mathbf{Sl}\left(  2,\mathbb{C}\right)  $
acts diagonally by
\[
g:\left(  \mathsf{A},\mathsf{B}\right)  \mapsto\left(  g\mathsf{A}%
g^{-1},g\mathsf{B}g^{-1}\right)  .
\]
Let $J:\ \left(  \mathsf{A},\mathsf{B}\right)  \mapsto\lbrack\mathsf{A}%
,\mathsf{B}]\ $be the momentum map$\ $on $X;$ the constraint locus is the cone%
\begin{equation}
Y=\{\left(  \mathsf{A},\mathsf{B}\right)  :b_{3}\left(  a_{1}-a_{2}\right)
-a_{3}\left(  b_{1}-b_{2}\right)  =0,\ b_{4}\left(  a_{1}-a_{2}\right)
-a_{4}\left(  b_{1}-b_{2}\right)  =0\}. \label{15}%
\end{equation}
Condition (\ref{10}) is easy to check. The polynomials%
\[
\mathrm{\alpha}_{1}=\mathrm{tr}\mathsf{A},\ \mathrm{\alpha}_{2}=\det
\mathsf{A},\ \mathrm{\beta}_{1}=\mathrm{tr}\mathsf{B},\ \mathrm{\beta}%
_{2}=\det\mathsf{B},\ \mathrm{\gamma}=\mathrm{\mathrm{tr}}\mathsf{AB}%
\]
generate the algebra $\mathcal{A}_{X/G}\ $of invariant polynomials on $X$. The
reduced Poisson bracket equals%
\begin{align}
q_{\mathrm{red}}  &  =2\frac{\partial}{\partial\mathrm{\alpha}_{1}}\wedge
\frac{\partial}{\partial\mathrm{\beta}_{1}}+\mathrm{\beta}_{1}\frac{\partial
}{\partial\mathrm{\alpha}_{1}}\wedge\frac{\partial}{\partial\mathrm{\beta}%
_{2}}+\mathrm{\alpha}_{1}\frac{\partial}{\partial\mathrm{\alpha}_{2}}%
\wedge\frac{\partial}{\partial\mathrm{\beta}_{1}}+\gamma\frac{\partial
}{\partial\mathrm{\alpha}_{2}}\wedge\frac{\partial}{\partial\mathrm{\beta}%
_{2}}\label{Q}\\
&  +\left(  \mathrm{\alpha}_{1}\frac{\partial}{\partial\mathrm{\alpha}_{1}%
}-\mathrm{\beta}_{1}\frac{\partial}{\partial\mathrm{\beta}_{1}}%
+2\mathrm{\alpha}_{2}\frac{\partial}{\partial\mathrm{\alpha}_{2}%
}-2\mathrm{\beta}_{2}\frac{\partial}{\partial\mathrm{\beta}_{2}}\right)
\wedge\frac{\partial}{\partial\gamma}.\nonumber
\end{align}

\begin{proposition}
The\ algebra $\mathcal{A}_{Y/G}$ of invariant polynomials of algebra
restricted to $Y$ is isomorphic to $\mathcal{B}/\left(  \mathrm{\rho}\right)
,$ where $\mathcal{B}=\mathbb{R}\left[  \mathrm{\alpha}_{1}\mathrm{,\alpha
}_{2}\mathrm{,\beta}_{1}\mathrm{,\beta}_{2}\mathrm{,\gamma}\right]  $ and
\begin{align}
\mathrm{\rho}  &  =\mathrm{\gamma}^{2}\mathrm{-\alpha}_{1}\mathrm{\beta}%
_{1}\mathrm{\gamma+\alpha}_{2}\left(  \mathrm{\beta}_{1}^{2}\mathrm{-2\beta
}_{2}\right)  \mathrm{+\beta}_{2}\left(  \mathrm{\alpha}_{1}^{2}%
\mathrm{-2\alpha}_{2}\right)  =\left(  \mathrm{\gamma-}\frac{1}{2}%
\mathrm{\alpha}_{1}\mathrm{\beta}_{1}\right)  ^{2}-4\mathrm{\pi},\ \label{S}\\
\mathrm{\pi}  &  =\mathrm{\tilde{\alpha}\tilde{\beta}},\ \mathrm{\tilde
{\alpha}}\doteqdot\mathrm{\alpha}_{2}\mathrm{-}\frac{1}{2}\mathrm{\alpha}%
_{1}^{2},\ \mathrm{\tilde{\beta}}\doteqdot\mathrm{\beta}_{2}\mathrm{-}\frac
{1}{2}\mathrm{\beta}_{1}^{2}.\nonumber
\end{align}

\end{proposition}

\textit{Proof. }Check that $\mathrm{\rho}=0$ on $Y.$ For any pair $\left(
\mathsf{A},\mathsf{B}\right)  \in Y,$ there exists $g\in\mathbf{Sl}\left(
2,C\right)  $\ such that both matrices $g\mathsf{A}g^{-1}$ and $g\mathsf{B}%
g^{-1}$ have Jordan form. This is easy to prove by means of (\ref{15}). Let
$\left(  a_{1},a_{2}\right)  $ and $\left(  b_{1},b_{2}\right)  $ be its
diagonal elements, respectively. Then
\[
\mathrm{\alpha}_{1}=a_{1}+a_{2},...,\ \mathrm{\beta}_{2}=b_{1}b_{2}%
,\ \mathrm{\gamma}=a_{1}b_{1}+a_{2}b_{2}%
\]
and (\ref{S}) can be checked directly. It is easy to show that this equation
generates all algebraic relations.$\ \blacktriangleright$

It follows that the spectrum of the algebra $\mathcal{A}_{Y/G}$ is a two-fold
covering of $\mathbb{K}^{4}$ ramified over the discriminant set $\left\{
\mathrm{\pi}=0\right\}  .$

\begin{conclusion}
The singular symplectic reduction of the variety $\left(  X,\mathbf{O}\left(
2\right)  ,q\right)  \mathbb{\ }$is the singular hypersurface $X_{\mathrm{red}%
}=\left\{  \mathrm{\rho}=0\right\}  $with coordinate functions $\mathrm{\alpha
}_{1}\mathrm{,\alpha}_{2}\mathrm{,\beta}_{1}\mathrm{,\beta}_{2}\mathrm{,\gamma
}$ defined by (\ref{S}) with the Poisson bracket $q_{\mathrm{red}}$ as in
(\ref{Q}).
\end{conclusion}

Let $\mathcal{A}^{\ast}$ be the extension of the algebra $\mathcal{A}_{X/G}$
by means of the element $\mathrm{\pi}^{-1/4}$.

\begin{proposition}
Elements
\begin{align*}
\mathrm{a}_{1}  &  =\frac{1}{\sqrt{2}}\mathrm{\alpha}_{1},\ \mathrm{b}%
_{1}=\frac{1}{\sqrt{2}}\mathrm{\beta}_{1},\ \\
\mathrm{a}_{2}  &  =\frac{\mathrm{\tilde{\alpha}}}{\mathrm{\pi}^{1/4}%
},\ \mathrm{b}_{2}=\frac{\mathrm{\tilde{\beta}}}{\mathrm{\pi}^{1/4}}%
\end{align*}
of algebra $\mathcal{A}^{\ast}$ fulfil (\ref{2}) with $n=2.$
\end{proposition}

\textit{Proof. }Obviously $\left\{  \mathrm{a}_{1},\mathrm{b}_{1}\right\}
=1.$ We have
\begin{equation}
q\left(  \mathrm{\tilde{\alpha},\tilde{\beta}}\right)  =\mathrm{\gamma-}%
\frac{1}{2}\left\{  \mathrm{\alpha}_{2}\mathrm{,\beta}_{1}\right\}
\mathrm{\beta}_{1}=\mathrm{\gamma}-\mathrm{a}_{1}\mathrm{b}_{1}. \label{3}%
\end{equation}
By (\ref{S})%
\begin{equation}
\mathrm{\gamma}-\mathrm{a}_{1}\mathrm{b}_{1}=2\mathrm{\pi}^{1/2} \label{4}%
\end{equation}
on $X_{\mathrm{red}}$ hence%
\[
q\left(  \mathrm{a}_{2},\mathrm{b}_{2}\right)  =\left\{  \frac{\mathrm{\tilde
{\alpha}}}{\mathrm{\pi}^{1/4}},\ \frac{\mathrm{\tilde{\beta}}}{\mathrm{\pi
}^{1/4}}\right\}  =\frac{1}{2\mathrm{\pi}^{1/2}}q\left(  \mathrm{\tilde
{\alpha},\tilde{\beta}}\right)  =1.\blacktriangleright
\]
This implies that the elements $\mathrm{a}_{k},\mathrm{b}_{k}\in
\mathcal{A}^{\ast}$ fulfil conditions (\ref{2}). By Proposition \ref{2n}%
\ bracket $q_{\mathrm{red}}$ admits a quantization by means of the GM series
with\ the hamiltonian fields \textrm{A}$_{k}=q\left(  \cdot,\mathrm{b}%
_{k}\right)  ,\ $\textrm{B}$_{k}=q\left(  \mathrm{a}_{k},\cdot\right)
.$\ These fields are well defined on since they vanish on the polynomial
$\mathrm{\rho}.$The explicit forms are%

\begin{align*}
\mathrm{A}_{1}  &  =\sqrt{2}\frac{\partial}{\partial\mathrm{\alpha}_{1}}%
+\frac{1}{\sqrt{2}}\mathrm{\alpha}_{1}\frac{\partial}{\partial\mathrm{\alpha
}_{2}}+\frac{1}{\sqrt{2}}\mathrm{\beta}_{1}\frac{\partial}{\partial
\mathrm{\gamma}},\ \\
\mathrm{B}_{1}  &  =\sqrt{2}\frac{\partial}{\partial\mathrm{\beta}_{1}}%
+\frac{1}{\sqrt{2}}\mathrm{\beta}_{1}\frac{\partial}{\partial\mathrm{\beta
}_{2}}+\frac{1}{\sqrt{2}}\mathrm{\alpha}_{1}\frac{\partial}{\partial
\mathrm{\gamma}},
\end{align*}%
\begin{align*}
\mathrm{A}_{2}  &  =\frac{3}{2}\mathrm{\pi}^{1/4}\frac{\partial}%
{\partial\mathrm{\alpha}_{2}}+\mathrm{\pi}^{-1/4}\mathrm{\tilde{\beta}}%
\frac{\partial}{\partial\mathrm{\gamma}}+\frac{1}{2}\mathrm{\pi}%
^{-3/4}\mathrm{\tilde{\beta}}^{2}\frac{\partial}{\partial\mathrm{\beta}_{2}%
},\\
\mathrm{B}_{2}  &  =\frac{3}{2}\mathrm{\pi}^{1/4}\frac{\partial}%
{\partial\mathrm{\beta}_{2}}+\mathrm{\pi}^{-1/4}\mathrm{\tilde{\alpha}}%
\frac{\partial}{\partial\mathrm{\gamma}}+\frac{1}{2}\mathrm{\pi}%
^{-3/4}\mathrm{\tilde{\alpha}}^{2}\frac{\partial}{\partial\mathrm{\alpha}_{2}%
}.\blacktriangleright
\end{align*}

\begin{conjecture}
Any term $\mathrm{Q}_{k},\ k=1,2,...$of the GM series is a bidifferential
operator of degree $\leq\left(  k,k\right)  $ with\ rational coefficients and
the denominator $\mathrm{\pi}^{k-1}.$
\end{conjecture}

This is obvious for $k=1$ since $\mathrm{Q}_{1}=q$. The direct calculation of
$\mathrm{Q}_{2}$ supports the conjecture.

\textbf{Other groups. }The above method works for the conjugacy action of the
orthogonal group $\mathbf{O}\left(  2\right)  $ on the space of pairs of real
symmetric $2\times2$ matrices as well for action of the unitary group
$\mathbf{SU}\left(  2\right)  $ on the space of pairs of Hermitian $2\times2$
matrices. The algebra of invariants is generated by the same five symmetric
polynomials. This bracket can be quantized in a similar way.

\section{Quantization of K3 surfaces}

K3 surfaces are topologically trivial Calabi-Yau 2-manifolds. A smooth variety
$X_{f}$ given in $\mathbb{CP}^{3}$ by a polynomial equation $f=0$ of degree 4
is a K3 surface. The Poisson bracket on $\mathcal{O}\left(  \mathbb{CP}%
^{3}\right)  /\left(  f\right)  $ is equal (up to a constant factor) to
$x_{0}^{-1}q_{0}$ on the chart $X_{0}=\left\{  x_{0}\neq0\right\}  $ where
\begin{equation}
q_{0}\left(  a,b\right)  =\det\left(
\begin{array}
[c]{ccc}%
\partial_{1}a & \partial_{2}a & \partial_{3}a\\
\partial_{1}b & \partial_{2}b & \partial_{3}b\\
\partial_{1}f & \partial_{2}f & \partial_{3}f
\end{array}
\right)  , \label{0}%
\end{equation}
$\partial_{i}=\partial/\partial x_{i},\ i=1,2,3$ and $x_{0},x_{1},x_{2},x_{3}$
are arbitrary homogeneous coordinates on\ $\mathbb{CP}^{3}\ $\cite{P1}in $CP$.
Below we consider two examples where Theorem \ref{WM} can be applied.

\textbf{I. }The variety\textbf{ }$X_{f}$ is a nonsingular K3 surface for
$f=1/4\left(  x_{0}^{4}+x_{1}^{4}+x_{2}^{4}+x_{3}^{4}\right)  .$ The canonical
Poisson bracket defined on $X_{f}$ is given by$\ q_{f}=x_{3}^{3}\partial
_{1}\wedge\partial_{2}+x_{1}^{3}\partial_{2}\wedge\partial_{3}+x_{2}%
^{3}\partial_{3}\wedge\partial_{1}\ $on the chart $X_{0}=\left\{
x_{0}=1\right\}  .$ We set $\mathrm{a}=\varphi\left(  x_{0},x_{3}\right)
x_{1},\mathrm{b}=\varphi\left(  x_{0},x_{3}\right)  x_{2}$ for an unknown
function $\varphi$ and solve the equation
\[
q_{f}\left(  \mathrm{a,b}\right)  =q_{f}\left(  \varphi\left(  x_{3}\right)
x_{1},\varphi\left(  x_{3}\right)  x_{2}\right)  =1.
\]
It is to check that $\varphi$ can be found in the form%
\[
\varphi^{2}\left(  x_{3}\right)  =\frac{2}{\sqrt{1+x_{3}^{4}}}\int^{x_{3}%
}\frac{\mathrm{d}y}{\sqrt{1+y^{4}}}.
\]

Singular surfaces in $\mathbb{CP}^{3}$ of degree 4.

\textbf{II. }Hypersurface $f=x_{0}x_{3}^{3}-x_{1}^{2}x_{2}^{2}$ has
singularities at four points where both terms $x_{0}x_{3}^{3},\ x_{1}^{2}%
x_{2}^{2}$ vanish. The bracket%
\[
q_{f}=3x_{0}x_{3}^{2}\partial_{1}\wedge\partial_{2}-2x_{1}x_{2}^{2}%
\partial_{2}\wedge\partial_{3}-2x_{1}^{2}x_{2}\partial_{3}\wedge\partial_{1}\
\]
is quantized on $X_{0}$ by\ the functions%
\[
\mathrm{a}=\frac{x_{2}}{x_{3}\sqrt{x_{0}}},\ \mathrm{b}=\frac{x_{1}}%
{x_{3}\sqrt{x_{0}}}.
\]
It is easy to check that $q_{f}\left(  \mathrm{a,b}\right)  =1$ which implies
that the hamiltonian fields
\begin{align*}
\mathrm{B}  &  =q_{f}\left(  \mathrm{a,}\cdot\right)  =-\frac{1}{\sqrt{x_{0}}%
}\left(  x_{0}x_{3}\partial_{1}+2x_{1}x_{2}^{3}x_{3}^{-2}\partial_{2}%
+2x_{1}x_{2}^{2}x_{3}^{-1}\partial_{3}\right)  ,\\
\mathrm{A}  &  =q_{f}\left(  \cdot,\mathrm{b}\right)  =-\frac{1}{\sqrt{x_{0}}%
}\left(  2x_{1}^{3}x_{2}x_{3}^{-2}\partial_{1}+x_{0}x_{3}\partial_{2}%
+2x_{1}^{2}x_{2}x_{3}^{-1}\partial_{3}\right)
\end{align*}
generate the quantization of the GM type.

\textbf{III. }If\textbf{ }$f=x_{0}^{2}x_{3}^{2}-x_{1}^{2}x_{2}^{2}$ we have
\[
q_{f}=2x_{0}^{2}x_{3}\partial_{1}\wedge\partial_{2}+2x_{1}x_{2}^{2}%
\partial_{2}\wedge\partial_{3}+2x_{1}^{2}x_{2}\partial_{3}\wedge\partial_{1}%
\]
and have $q_{f}\left(  \mathrm{a}_{1},\mathrm{b}_{1}\right)  =1$ if we take%
\[
\mathrm{a}=\frac{x_{1}}{2x_{0}\sqrt{x_{3}}},\ \mathrm{b}=\frac{x_{2}}%
{2x_{0}\sqrt{x_{3}}}.
\]

\textbf{IV. }For $f=x_{3}^{4}-x_{1}^{2}x_{2}^{2}$ we have$\ q_{f}\left(
\mathrm{a},\mathrm{b}\right)  =1$\ for the elements
\[
\mathrm{a}=\frac{x_{1}}{x_{3}\sqrt{x_{3}}},\ \mathrm{b}=-\frac{x_{2}}%
{x_{3}\sqrt{x_{3}}}.
\]

\end{document}